\documentclass[12pt]{iopart}
\usepackage{graphicx}
\usepackage{amssymb,latexsym}
\usepackage{amsfonts}
\usepackage{amssymb}
\usepackage{cancel}
\usepackage{color}
\input{colordvi.tex}

\begin{document}

\title[Compact stars with a small electric charge]{Compact stars with
a small electric charge: the limiting radius to mass relation and the
maximum mass for incompressible matter}

\author{Jos\'e P. S. Lemos}
\address{Centro Multidisciplinar de Astrof\'isica - CENTRA,
Departamento de F\'{\i}sica, Instituto Superior T\'ecnico - IST,
Universidade de Lisboa - UL, Av. Rovisco Pais 1, 1049-001
Lisboa, Portugal; joselemos@ist.utl.pt}

\author{Francisco J. Lopes}
\address{Centro Multidisciplinar de Astrof\'isica - CENTRA,
Departamento de F\'{\i}sica, Instituto Superior T\'ecnico - IST,
Universidade de Lisboa - UL, Av. Rovisco Pais 1, 1049-001
Lisboa, Portugal; franciscojoaolopes@ist.utl.pt}

\author{Gon\c calo Quinta}
\address{Centro Multidisciplinar de Astrof\'isica - CENTRA,
Departamento de F\'{\i}sica, Instituto Superior T\'ecnico - IST,
Universidade de Lisboa - UL, Av. Rovisco Pais 1, 1049-001
Lisboa, Portugal;  goncalo.quinta@ist.utl.pt}

\author{Vilson T. Zanchin}
\address{Centro
de Ci\^encias Naturais e Humanas, Universidade Federal do ABC, Rua
Santa Ad\'elia 166, 09210-170 Santo Andr\'e, SP, Brazil; 
zanchin@ufabc.edu.br}

\newpage
\begin{abstract}
One of the stiffest equations of state for matter in a compact star is
constant energy density and this generates the interior Schwarzschild
radius to mass relation and the Misner maximum mass for
relativistic compact stars. If dark matter
populates the interior of stars, and this matter is supersymmetric or
of some other type, some of it possessing a
tiny electric charge, there is the possibility that 
highly compact stars can trap a small
but non-negligible electric charge.  In this case the radius to mass
relation for such compact stars should get modifications.  We use an
analytical scheme to investigate the limiting radius to mass relation
and the maximum mass of relativistic stars made of 
an incompressible fluid with a small electric
charge. The investigation is carried out by
using the hydrostatic equilibrium equation, i.e., the
Tolman-Oppenheimer-Volkoff (TOV) equation, together with the other
equations of structure, with the further hypothesis that the charge
distribution is proportional to the energy density. The approach
relies on Volkoff and Misner's method to solve the TOV equation. For
zero charge one gets the interior Schwarzschild limit, and supposing
incompressible boson or fermion matter with constituents with masses
of the order of the neutron mass one gets that the maximum mass is the
Misner mass.  For a small electric charge, our analytical
approximating scheme valid in first order in the star's electric
charge, shows that the maximum mass increases relatively to the
uncharged case, whereas the minimum possible radius decreases, an
expected effect since the new field is repulsive aiding the pressure
to sustain the star against gravitational collapse.

\end{abstract}

\maketitle

\section{Introduction}
\label{sec-introd}

Compact stars and their properties have been a theme of great
relevance on several grounds. 
Chandrasekhar's celebrated work \cite{chandra1} on the
maximum mass for white dwarfs advanced the way to the understanding of
the nature and structure of compact stars. By using a
cold equation of
state in which the degeneracy electron pressure is the most relevant
form of pressure 
for the support of a white dwarf against gravitational collapse, a
radius-mass relation for these stars was deduced, from the
non-relativistic electron regime in relatively large white dwarfs up
to the relativistic electron regime in the most compact stars. 
He found that as the radius of
the star approached zero
the mass would go to a maximum
value of $1.44\,M_\odot$, where $M_\odot$
is the sun's mass. This is the Chandrasekhar limit. It uses
Newtonian gravitation.  Landau \cite{landau} through heuristic
arguments found that the mass limit for white dwarfs could be written
as $M\sim M_{\rm pl}^3/m_{n}^2$ where $m_{n}$ is the neutron
or the proton mass and $M_{\rm pl}$ is the Planck mass, $M_{\rm
pl}=\sqrt{\hbar c/G}$, with $\hbar$ being the Planck constant, 
$G$ the Newton's constant of gravitation, and $c$
the velocity of light, 
or, setting units such that $G=1$ and $c=1$, which 
we do from now on, one has  $M_{\rm
pl}=\sqrt{\hbar}$.
Putting in the numerical values for $M_{\rm pl}$
and $m_n$, the mass $M$ of the star is
about the Chandrasekhar mass limit $M\sim 1\,\,M_\odot$. Landau further
deduced that the stars should have a radius of about $\lambda_{e}\,
M_{\rm pl}/m_{n}$, where $\lambda_{e}$ is the electron's
Compton wavelength, $\lambda_{e}=\hbar /m_{e}$, $m_{e}$
being the electron's mass, giving a radius 
of the order of 5000$\,$km.  He also found that
there was another regime in which the star is composed of neutrons,
supported by the degeneracy pressure of these particles, and has a
maximum mass of about $M\sim M_{\rm pl}^3/m_{n}^2$.  These neutron
stars are much more compact with a radius $\lambda_{n}\, M_{\rm
pl}/m_{n}$, $\lambda_{n}$ being the neutron's Compton
wavelength, giving about 10 km.  For objects with a radius tending to
zero one should use general relativity, rather than Newtonian
gravitation. In general relativity a compact star can be defined
neatly as a star that has a geometrical mass $M$
(or, $GM/c^2$ if one restores $G$ and $c$) somehow
comparable to its radius $R$,
i.e., $R/M\sim a$, with $a$ a number not much bigger
than 1, say of order of 10 or less.
Whereas for an extended star like the
Sun $R/M_\odot\sim 5\times 10^5$, one has for a white dwarf $R/M_\odot
\sim 3
\times 10^3$, and for a neutron star $R/M_\odot\sim 6$,
showing that the latter is really compact.  On a general relativistic
basis, Oppenheimer and Volkoff
\cite{oppievolkoff} worked out further
the nature and structure of neutron stars.  Setting up a stiff
equation of state for stars matter made of neutrons they found roughly
the results of Landau 
\cite{landau}, namely, the mass limit is about $1\,\,M_\odot$
and $R/M_\odot\sim 6$.  This limit is called the
Landau-Oppenheimer-Volkoff limit.  Improvements have been made on
these limits.  Using a cold equation of state valid in the full range
of highly compressed matter, the Harrison-Wheeler equation of state,
the full set of equilibria in an $R$-$M$ relation were found, in
particular, the two maxima masses corresponding to the Chandrasekhar
and Landau-Oppenheimer-Volkoff limits appear naturally \cite{htww}.
See also \cite{glendenningbook} for further discussion on compact
stars
and \cite{rezmaart}
for issues of their
stability. These mass limits, as seen in the context of general
relativity, appear because at these stages the energy associated to
the pressure is so strong that its gravitating effect overwhelms the
self support effect.

The properties of these compact
stars get modified if
either the constituent material is altered 
or an alternative theory of
gravitation is used.  
The constituent material 
can be altered by 
introducing or adding new matter 
with different properties
or even adding a new field to the star. 
It is also
well known that the introduction of new matter fields can be mimicked
by modifications of the gravitational field.  
One example, even in
Newtonian gravitation, is that the effects of the dark matter can be
mimicked by modifying the gravitational field, e.g., in the MOND
theories \cite{milg1,milg2,milg3}.  The same type of choice holds
true in tensor theories of gravitation, since one can pass the excess
of the gravitational field present in the left-hand side of the
Einstein equations to their right hand side giving an effective
energy-momentum tensor in a form of a new field, e.g., see
\cite{famaey}. 
If this dark matter
inhabits the core of stars it can
imprint onto  compact stars new properties
\cite{kouv1,kouv2}. On the other hand, the structure of stars like
neutron stars in alternative theories of gravity has been analyzed 
in 
an Einstein-Dilaton-Gauss-Bonnet gravity \cite{pani},
in
Eddington-like theories 
\cite{harko} with the conclusion that more massive stars than in
general relativity can form, 
in
a quadratic gravity theory
\cite{delidu,yazadjiev}
and in  $f(R)$ theories
\cite{cheoun,astashenok}.
Stars and compact stars in braneworlds
have also been studied in \cite{deruelle,germaart,hladík,castro}.

In contrast to the fermion stars mentioned above, 
there are
stars made of bosons, the boson stars, that can have
a wide range of mass limits, namely, $M\sim M_{\rm pl}^2/m_{b}$,
$M\sim M_{\rm pl}^3/m_{b}^2$, or $M\sim M_{\rm pl}^4/m_{b}^3$, where
$m_{b}$ is the mass of the boson that makes up the star
\cite{kaup,rufbonazz,lee}, see \cite{jetzer,liebpal} for reviews.
These stars could have been formed in the beginning of the universe
from the primordial gravitational collapse of the boson particles and
have been proposed as alternatives to the usual compact objects
\cite{liebpal}, such as neutron stars and black holes, and also as
part of the dark matter \cite{tamaki}. These stars can, in principle,
be detected \cite{dabrowski,yuan,palenzuela}
and a way to tell the 
the difference between fermionic and
bosonic dark matter has been proposed in 
\cite{hannestad}.

Now, the first compact star ever displayed in its 
full structure was a general
relativistic star with a very stiff equation of state, a star made
of an incompressible perfect fluid, i.e., $\rho(r)={\rm constant}$, and
isotropic pressure (where $\rho(r)$ is the energy density at the
radius $r$) \cite{incomschwarzschild}.  This interior Schwarzschild
star solution is spherically symmetric and has a
vacuum exterior.  
An incompressible equation of state is interesting from
various aspects, since one can extract clean results and it also provides
compactness limits.  
Furthermore, this incompressible
fluid applies to both fermion and boson particles, as long as the
fluid has an 
ultra-stiff
equation of state that can be approximated by 
an incompressible state. 
The speed of sound $(dp/d\rho)^{1/2}$
in such a fluid, is infinite,
which is not
allowed relativistically. Notwithstanding, it is believed that the
interiors of dense stars, such as neutron stars, 
are of uniform density, or nearly so, and thus
this simple case is of practical interest.
Schwarzschild found
that there was a limit, when the central pressure $p_c$ goes to
infinity and that the star's radius to the mass limit is
$R/M=9/4=2.25$, indeed a very compact star
\cite{incomschwarzschild}.  Volkoff \cite{volkoff}
and Misner \cite{misner} rederived the Schwarzschild interior limit of
$R/M=9/4$ using the convenient Tolman-Oppenheimer-Volkoff (TOV)
equation, a differential equation for the pressure profile as a
function of the other quantities \cite{oppievolkoff}, see also
\cite{tolman}. In addition, Misner \cite{misner} even found a
maximum mass for a given density of the incompressible fluid, the
Misner mass.  The Schwarzschild limit yields, for the same mass, a
radius that is well below the radius of a neutron star, and also
yields the most possible compact boson star. An incompressible star
whose radius is below this limit will in principle collapse into a
singularity leaving a black hole to the exterior.

One can ask if the Schwarzschild limit can be modified, allowing for
instance a lower $R/M$ relation.  As mentioned above, one way is to
have some kind of different matter or a new field in the star,
or an alternative theory of gravitation.  
Rather than introducing an alternative theory of
gravitation one can opt for a fluid with non-isotropic pressure
\cite{makharko}, for instance. 
Here we select to study the case in which we add a matter
field to the existent matter. We consider matter with a small electric
charge, introducing thus an additional electric field in addition to
the usual matter and gravitational fields.  This addition of an
electric charge and an electric field to the Schwarzschild
incompressible matter configurations brings insight to the
configurations overall structure in more complex situations and its
study in stars mimics other fields and possible alterations in the
gravitational field.

Spheres of electric charge tend to be unstable. Since like charges
repel each other, if in one way or another
a charged cloud forms  it
will blow out at once due to the electrical repulsion. However,
gravitation is universal and attractive.  
For a sphere with a small mass the electric repulsion is
stronger than the gravitational pull and the matter suffers an
expansion. On the other hand,
for a highly massive sphere
with a small quantity of charge the gravitational pull on the charged
matter can overcome the electric repulsion and thus the system
collapses. In between the two 
situations, one can have an equilibrium situation when the
gravitational pull and the electric repulsion balance each other and
the sphere stays static. In general, there is also matter pressure
which adds to the balance of forces. For an extended star the pressure
acts to counterbalance the gravitational field, but for highly compact
stars the pressure acts as an energy source that adds to the star's
mass and energy increasing in an overwhelming fashion the
gravitational pull against its own pressure push.  Thus, in the
balance between the gravitational, electric and pressure fields, in
highly compact stars electric charge can remain trapped within the
stars.  In case the compact stars supports some electric charge, their
structure and properties are modified, and in particular, the
radius-mass relations for the corresponding stars should change.

One important quantity that gives a 
measure of how much electric charge a star can
support is the ratio of the mass $m$ to the charge $q$ of the main
fundamental constituents of the star. For
normal matter the net electric matter in a star is utterly negligible as
the ratio of the proton mass $m_{p}$ to the proton charge $e$ is
$m_{p}/e=10^{-18}$, giving thus $Q/M\simeq
(m_{p}/e)^2\simeq10^{-36}$, where $Q$ is the star's total
charge \cite{glendenningbook}
(see also \cite{iorio}).  However, stars
can contain some dark matter in their interior, and of the several
dark matter fluid candidates some could be electrically charged.
Indeed, natural candidates to compose the dark matter are
supersymmetric particles.  The lightest supersymmetric particles that
make the bulk of dark matter should be neutral, one
possible candidate is the
neutralino \cite{byrne}, however, some of these particles could be
electrically charged.  The mass $m$ to charge $q$ ratio of these
supersymmetric particles are much higher than the baryonic mass to
charge ratio, indeed current supergravity theories indicate that some
particles can have a ratio of one.  For a $m/q\sim 0.1-0.3$
one has $Q/M\simeq 0.01-0.1$, a small but non-negligible
electric charge. Thus, if dark matter populates the interior of stars,
and some of it is made of electrically charged particles there is the
possibility that stars have some electric charge.

That electric charge can influence the structure of a compact star was
proposed earlier by 
Bekenstein
\cite{bekenstein} who wrote the appropriate TOV
equation.  Some electric compact configurations with an incompressible
equation of state for the matter were studied numerically and the
corresponding generalized Schwarzschild limit, i.e., 
central pressure going
to infinity in these configurations, was analyzed
\cite{defeliceyu,defelicesiming,annroth,arbalemoszanchin}.  Other
equations of state for electrically charged matter, such as polytropic
equations were used in
\cite{raymalheirolemoszanchin,ghezzi2005,alz-poli-qbh}, where star
configurations and their structure were studied and the Schwarzschild
electric limit for the given equation of state and for a given charge
was considered. In particular, in \cite{raymalheirolemoszanchin}
it was argued that upon gravitational 
collapse a star could retain, significantly, 
part of its electric charge.
This electric charge excess could remain 
trapped in the final configuration, be it 
a highly compact star
or a black hole.
Other interesting equations of state were proposed
and studied in \cite{guilfoyle,lemosezanchin_QBH_pressure}, and in
\cite{lemosweinberg} where electrically charged dust was studied.
Electrically charged boson stars have been also studied and their
properties analyzed \cite{jetzer2,brihaye}.  
Bounds on the radius to
mass relation for uncharged 
stars have been put forward in \cite{buchdahl}, see 
also \cite{bondi,islam1,islam2,mv}.
Bounds on the radius to
mass relation for
charged stars have been set in
\cite{andreasson_charged}, see also
\cite{sim,bohmerharko,giulianirothman}.  It is also worth mentioning
some work on charged Newtonian stars. That compact stars could exist
was even noticed in the 18th century. A subset of these Newtonian
compact stars are the dark stars of Mitchell, also mentioned later by
Laplace, see \cite{isra}. The Chandrasekhar white dwarfs of very small
radius, including the one with zero radius that gives the
Chandrasekhar maximum mass, are also Newtonian compact stars. Indeed
the stars that have very small radii, from the gravitational radius to
zero radius, provide instances of the dark stars of Mitchell and
Laplace. Of course these stars cannot exist in nature as for strong
gravitational fields Newtonian gravitation is invalid.  That compact
Newtonian stars could be electrified was raised in \cite{bonnor1} and
further developed in \cite{lemoszanchinbon}.  Turning the table
around, the real analogue of dark stars are the quasiblack holes
considered in, e.g., \cite{lemosezanchin_QBH_pressure}.

In order to understand the effects of 
a small electric charge on a star, and in 
particular, on the interior
Schwarzschild limit, we use an analytical scheme and investigate the
limiting radius to mass relation and the maximum mass of relativistic
compact stars made of an electrically charged incompressible fluid. The
investigation is carried out using the hydrostatic equilibrium
equation, i.e., the TOV equation, and the other structure equations,
with the further hypothesis that the charge distribution is
proportional to the energy density. The approach relies on Volkoff and
Misner's method \cite{volkoff,misner} to solve the TOV equation. For
zero charge one gets the interior Schwarzschild limit and under
certain assumptions one gets the Misner mass. Our analysis for stars
with a small electric charge shows that the maximum mass increases
relatively to the uncharged case, whereas the minimum possible radius
decreases, an expected effect since the new field is repulsive aiding
the pressure to sustain the star against gravitational collapse.

The paper is organized as follows. In Sec.~\ref{sec-basicequations} we
give the general relativistic equations, the equations of structure
for a static spherically symmetric configuration, the equations of
state for energy density and charge density, and discuss the 
exterior spacetime and the boundary
conditions.  In Sec.~\ref{zerocharge} we review the pure, uncharged,
interior Schwarzschild limit using the Volkoff and Misner's formalism
to set the nomenclature. We also give the Misner mass.  In
Sec.~\ref{smallcharge} we study analytically the interior electric
Schwarzschild limit using the Volkoff and Misner's formalism
and give the electric correction to the Misner mass. 
 In Sec.~\ref{sec-conclusion} we conclude.  In
the appendix we derive some equations necessary in our study.

\section{The full set of equations
and boundary conditions}
\label{sec-basicequations}

\subsection{Basic general relativistic equations}
\label{sec-basicequations2}

We are interested in analyzing  highly compacted
charged spheres as described by
the Einstein-Maxwell equations
with charged matter. We set $G=1$ and
$c=1$. The field equations are
\begin{eqnarray}
&& G_{\mu\nu}=8\pi T_{\mu\nu},\label{eqs de einstein}\\
&& \nabla_{\nu}F^{\mu\nu}=4\pi J^{\mu}, \label{eqs de maxwell}
\end{eqnarray} 
where Greek indices are spacetime indices running from $0$ to $3$,
with $0$ being a time index.  The Einstein tensor $G_{\mu\nu}$ 
is defined as 
$G_{\mu\nu} =R_{\mu\nu} -\frac{1}{2}g_{\mu\nu} R$,
where $R_{\mu\nu}$ is the Ricci tensor $R_{\mu\nu}$, 
$g_{\mu\nu}$is the metric tensor, and $R$  the Ricci scalar.  
The 
Faraday-Maxwell tensor $F_{\mu\nu}$
is defined in terms of an 
electromagnetic four-potential $A_\mu$ by
$F_{\mu\nu}=\partial_\mu\,A_\nu-
\partial_\nu\,A_\mu$.
Equation~(\ref{eqs de einstein}) is the Einstein equation,
stating the relation between the Einstein tensor
and the energy-momentum tensor $T_{\mu\nu}$.
$T_{\mu\nu}$ is written here
as a sum of two terms,
\begin{equation}\label{tensor1 de energia momento}
T_{\mu\nu}=E_{\mu\nu}+M_{\mu\nu}\,.
\end{equation}
$E_{\mu\nu}$ is 
the electromagnetic energy-momentum tensor, which is given in terms of
the Faraday-Maxwell tensor $F_{\mu\nu}$ by the relation
\begin{equation}\label{tensor de energia momento}
E_{\mu\nu}=\frac{1}{4\pi}\left(
F_{\mu}\hspace{0.1mm}^{\gamma}F_{\nu\gamma}-
\frac{1}{4}g_{\mu\nu}
F_{\gamma\beta}F^{\gamma\beta}\right)\,.
\end{equation}
$M_{\mu\nu}$
represents the matter energy-momentum tensor
and we assume to be 
the energy-momentum tensor of a perfect fluid, namely,
\begin{equation}
M_{\mu\nu}=(\rho+p)U_{\mu}U_{\nu}+ pg_{\mu\nu},
\end{equation}
with $\rho$ and $p$ being the energy density and the pressure of the
fluid, respectively, and $U_\mu$ is the fluid four-velocity.  
Equation~(\ref{eqs de maxwell}) is the Maxwell equation,
stating the proportionality between the covariant 
derivative $\nabla_\nu$ of the Faraday-Maxwell tensor $F_{\mu\nu}$
and the electromagnetic
four-current $J_\mu$. For a charged fluid, this current 
is given in terms of the
electric charge density $\rho_e$ by
\begin{equation}
J^{\mu}=\rho_{e}U^{\mu}\,.
\end{equation}
The other Maxwell equation $\nabla_{[\alpha}F_{\beta\gamma]}=0$,
where $[...]$ means antisymmetrization, 
is automatically satisfied.

\subsection{Equations of structure}
\label{sec-equilibriumequations}

The line element 
for a static spherically symmetric spacetime 
is of the form
\begin{equation}\label{geral_metric}
ds^2=-B(r)\,dt^{2}+A(r)\,dr^2+r^2\left(d\theta^2+\sin^{2}\theta
d\phi^{2}\right),
\end{equation}
where $t,\, r,\, \theta$ e $\phi$ are the usual Schwarzschild-like
coordinates, and the metric potentials $A(r)$ and $B(r)$
are functions of the radial coordinate $r$ only.
The assumed spherical symmetry of the spacetime implies that the only
nonzero components of a purely electrical
Faraday-Maxwell tensor $F^{\mu\nu}$ are
$F^{tr}$ and $F^{rt}$
with
$F^{tr}=-F^{rt}$ and where $F^{tr}$ is a function of the radial
coordinate $r$ alone, $F^{tr}=F^{tr}(r)$. The other components of
$F^{\mu\nu}$ are identically zero. It is advantageous to 
define 
the total electric charge $q(r)$ inside a spherical
surface labeled by the radial coordinate whose value is $r$
by
\begin{equation}\label{geral_maxwell}
q(r)=F^{tr} r^2\sqrt{A(r)\,B(r)}\,.
\end{equation}
I.e., one swaps $F^{tr}$ for $q(r)$. 
It is also
opportune to define a new quantity $m(r)$ in such a way that
\begin{equation}\label{funcion metrica}
\frac{1}{A(r)}=1-\frac{2m(r)}{r}+\frac{q^{2}(r)}{r^{2}}.
\end{equation}
I.e., one swaps $A(r)$ for $m(r)$.
The new function $m(r)$ represents
the gravitational mass inside the sphere of radial coordinate $r$. 

One of the Einstein equations can be substituted by the
contracted Bianchi identity
$\nabla_{\mu}T^{\mu\nu}=0$, which gives
\begin{equation}\label{conservacion2}
\frac{dB(r)}{dr}=\frac{B(r)}{p(r)+\rho(r)}\left[\frac{q(r)}{2\pi
r^{4}}\frac{dq(r)}{dr}-2\frac{dp(r)}{dr}\right]\,,
\end{equation}
a differential equation for $B$, $q$, and $p$.
Einstein equations
also give a differential equation for $B(r)$ alone, i.e.,
\begin{equation}
\left({1-\frac{2m(r)}{r}+\frac{q^{2}(r)}{r^{2}}}\right)\left[1+
\frac{r}{B(r)}\frac{dB(r)}{dr}\right]
= {1}
+ 8\pi r^2\left[p(r)-\frac{q^{2}(r)}{8\pi r^{4}}\right].
\label{G11final12}
\end{equation}
Now, we are ready to write the
other three equations in a form
we want to use.
One finds that another of Einstein equations gives
a differential equation for $m(r)$, i.e., 
\begin{equation}\label{continuidaddamassa}
\frac{dm(r)}{dr}=4\pi\rho(r) r^{2}+\frac{q(r)}{r}
\frac{dq(r)}{dr}\,.
\end{equation}
Since $m(r)$
represents the gravitational mass inside the sphere of radial
coordinate $r$, Eq.~(\ref{continuidaddamassa}) represents then the
energy conservation as measured in the star's frame.
The only non-vanishing
component of the Maxwell equation (\ref{eqs de maxwell}) is
given by 
\begin{equation}
\label{continuidadedacarga}
\frac{dq(r)}{dr}=4\pi\rho_{e}(r)\,r^{2} 
\sqrt{{1-\frac{2m(r)}{r}+\frac{q^{2}(r)}{r^{2}}}},
\end{equation}
Finally, replacing Eq.~(\ref{continuidadedacarga}) and the
conservation equation (\ref{conservacion2}) into 
Eq.~(\ref{G11final12}) it yields
\begin{eqnarray}\label{tov}
\frac{dp}{dr}= &&-(p+\rho)\frac{\left(4\pi pr+m/r^{2}
-q^{2}/r^{3}\right)}{\left(1-2m/r
+q^{2}/r^{2}\right)}+\rho_{e}\,\frac{q/r^2}{ \sqrt{1-2m/r
+q^{2}/r^{2}}}\,,
\end{eqnarray}
where to simplify the notation we 
have dropped the functional dependence, i.e., 
$m(r)=m$, $q(r)=q$, $\rho(r)=\rho$, 
$p(r)=p$, and $\rho_{e}(r)=\rho_{e}$. 
Eq.~(\ref{tov}) is the TOV
equation 
modified by the inclusion of electric
charge \cite{bekenstein} (see also \cite{alz-poli-qbh}).
The system of equations 
(\ref{G11final12})-(\ref{tov}) is 
the system we were looking for.
We need now
to specify the equation of state and 
the equation for the charge density profile.

\subsection{Equation of state and the charge density profile}
\label{EOS_charge_DP}

In the present model there are six unknown functions: $B(r)$, $m(r)$,
$q(r)$, $\rho(r)$, $p(r)$, and $\rho_{e}(r)$; and just four equations:
Eqs.~(\ref{G11final12}), (\ref{continuidaddamassa}),
(\ref{continuidadedacarga}), and (\ref{tov}).
Additional relations are obtained
from a model for the cold fluid, which
furnishes relations among the pressure and the energy density.  For an
electrically charged fluid, a relation defining the electric charge
distribution is also needed.

Here we assume an incompressible fluid, i.e.,
\begin{equation}\label{density_0}
\rho(r)={\rm constant}\,.
\end{equation}
So the energy density is constant along the whole star.

Following \cite{raymalheirolemoszanchin} (see also
\cite{alz-poli-qbh}), we assume
a charge density proportional to the energy density,
\begin{equation}\label{densicarga_densimassa}
\rho_e=\alpha\rho\,,
\end{equation}
where, in geometric units, $\alpha$ is a dimensionless constant which
we call the charge fraction. The charge density along the whole star
is thus constant as well.
Other 
equations for the charge
distribution could be considered, as 
more charge concentration on the core, or more
charge on the outer layers, see , e.g., 
\cite{defeliceyu,defelicesiming,annroth}.
An equation for the charge density as the one given 
in (\ref{densicarga_densimassa}) should be commented.  
In principle, the permittivity $\varepsilon$ 
of such a medium cannot
be equal to the vacuum 
permittivity $\varepsilon_{\rm vac}=1$. Such 
a $\varepsilon\neq\varepsilon_{\rm vac}$ 
certainly has influence on the 
electrostatic equation (\ref{continuidadedacarga}) and also 
possibly on the 
stress-energy tensor $E_{\mu\nu}$ 
of the electromagnetic field given in 
Eq.~(\ref{tensor de energia momento}).
Here we consider that the medium is such that 
its response to an applied electric 
field is low enough so that one can consider 
that it has a permittivity equal to 
the vacuum permitivity. In a more detailed
account permittivity effects should be considered.

We have now four equations: Eqs.~(\ref{G11final12}), 
(\ref{continuidaddamassa}), (\ref{continuidadedacarga}),
and
(\ref{tov}); and four unknowns: $B(r)$, $m(r)$,
$q(r)$, and $p(r)$, as $\rho$ and
$\rho_e$ are given in 
(\ref{density_0}) and (\ref{densicarga_densimassa}), respectively.
The resulting set of
equations constitute the complete set of structure equations which,
with some appropriate boundary conditions, can be solved
simultaneously. We are not going to solve it
(see 
\cite{arbalemoszanchin}). 
Here we use this system of equations to find
the Schwarzschild interior limit for the small
charge case.

\subsection{The exterior vacuum 
region to the star and the boundary conditions}
\label{BC}

The conditions at the center of the star are that 
$m(r=0)=0$, $q(r=0)=0$, and $A(r=0)=1$
to avoid any type of singularities,
and that 
 $p(r=0) = p_c$, $\rho(r=0)=\rho_{c}$, and 
$\rho_e(r=0)=\rho_{ec}$,
where $p_c$ is the central
pressure, $\rho_{c}$ 
is the central energy
density, and 
$\rho_{ec}$ is the central charge distribution,
the two latter having the same 
constant values throughout the
star (see Eqs.~(\ref{density_0}) and
\ref{densicarga_densimassa}).

The interior solution is matched 
at the surface to the exterior
Reissner-Nordstr\"om spacetime, with metric given by
\begin{equation}
ds^2 = - F(r)\, dT^2 + \frac{dr^2}{F(r)}+ r^2
\left(d\theta^2+\sin^{2}\theta
d\phi^{2}\right)\,,
\label{rnords1}
\end{equation}
where
\begin{equation}
F(r) = 1 -2M/r + Q^2/r^2\, ,
\label{rnords2}
\end{equation}
with the outer time $T$ being proportional to the inner time $t$, 
and $M$
and $Q$ being the total mass and the total charge of the star,
respectively.

At the surface of the star one has a vanishing pressure, i.e.,
$p(r=R)=0$.  The boundary conditions at the surface of the star are
then $B(R)=1/A(R)=F(R)$, $m(R)=M$, $q(R)=Q$, besides $p(R)=0$.

An important quantity for the exterior metric is the gravitational or
horizon radius $r_+$ of the configuration. The Reissner-Nordstr\"om
metric, given through Eqs.~(\ref{rnords1})-(\ref{rnords2}), 
has 
\begin{equation} 
r_+ = M +
\sqrt{M^2 - Q^2} \,.
\label{externalgravradius} 
\end{equation}
as the solution for its own gravitational radius.

\section{The interior Schwarzschild limit: The zero charge case}
\label{zerocharge}

\subsection{Equations}

Before we treat the small charge case analytically, we consider the
exact Schwarzschild interior solution as given by Volkoff 
\cite{volkoff} and displayed later in Misner's lectures
\cite{misner}. For this we put $q=0$ in
Eqs.~(\ref{G11final12})-(\ref{tov}).
Equation (\ref{G11final12}) is of no direct interest here,
Eq.~(\ref{continuidaddamassa}) gives
\begin{equation}\label{continuidaddamassauncharged}
\frac{dm(r)}{dr}=4\pi\rho(r) r^{2}\,,
\end{equation}
Eq.~(\ref{continuidadedacarga}) is trivially satisfied
in this case, and finally, Eq.~(\ref{tov})
simplifies to 
\begin{equation}\label{tovuncharged}
\frac{dp}{dr}=-(p+\rho)\frac{4\pi pr+m/r^{2}
}{1-2m/r}
\,.
\end{equation}
Since, by equation (\ref{density_0}), the density is constant we can
integrate equation (\ref{continuidaddamassauncharged}) obtaining
\begin{equation} \label{mass0}
m(r) = \frac{4}{3} \pi \rho\, r^3\,,\quad 0\leq r\leq R\,,
\end{equation}
where $R$ is the star radius and we have imposed
that there is no point mass in the center.
Defining a characteristic length $R_c$ as
\begin{equation}
R_c^2=\frac{3}{8 \pi \rho},
\label{length}
\end{equation}
we can rewrite the mass function, Eq.~(\ref{mass0}), as
\begin{equation} \label{mass0asR}
m(r) = \frac12\frac{r^3}{R_c^2} \,,\quad 0\leq r\leq R\,.
\end{equation}
Interchanging $\rho$ and $R_c$ as necessary and noting that 
$2\,\rho\,R_c^2=\frac{3}{4\pi}$ we get from 
Eq.~(\ref{tovuncharged}), 
\begin{equation}\label{tovuncharged1}
\frac{dp}{dr}=-\frac{(p+\rho)(3p+\rho)}{\rho}\,\frac{1}{2R_c^2}
\frac{r}{1-r^2/R_c^2} 
\,.
\end{equation}

\subsection{The interior Schwarzschild limit: The $R$ and 
$M$ relation and the minimum radius}

Equation (\ref{tovuncharged1}) is separable and can be integrated as
\begin{equation}\label{tovuncharged2}
\int dp\, \frac{\rho }{(\rho + p)(\rho + 3p)}=-\frac12
\int dr\,
\frac{r }{R_c^2}
\frac{1}{1-r^2/R_c^2} 
\,, 
\end{equation}
with the boundary condition that the surface of the star
$R$
has zero pressure, i.e., 
\begin{equation}\label{bc1tovuncharged2}
p(R)=0
\,.
\end{equation}
Defining a new radial coordinate $\chi$ by
\begin{equation}
r = R_c \sin \chi\,,
\label{chi}
\end{equation}
Eq.~(\ref{tovuncharged2}) can be put in the form
\begin{equation}\label{tovuncharged21}
\int dp\, \frac{\rho }{(\rho + p)(\rho + 3p)}=-\frac12
\int\, d\,(\ln \cos \chi)\,.
\end{equation}
subjected to the  boundary condition
\begin{equation}
p(\chi_s)=0\,,
\label{cfp}
\end{equation}
where $\chi_s$ is given through 
\begin{equation}
R = R_c\,\sin \chi_s\,.
\label{chis}
\end{equation}
Integrating Eq.~(\ref{tovuncharged21}), subjected to 
the boundary condition (\ref{cfp}), yields the pressure
\begin{equation}
p = \rho \frac{\cos \chi - 
\cos \chi_s}{3 \cos \chi_s - \cos \chi}.
\label{tovunchargedsol}
\end{equation}
The central pressure, $p_c$ is the pressure computed at 
zero radius $r=0$, i.e., 
$\chi=0$, so that
\begin{equation}
p_c = \rho \frac{1 - \cos \chi_s}{3 \cos \chi_s - 1}\,.
\label{pcuncharged}
\end{equation}
This blows up, 
\begin{equation}
p_c \rightarrow \infty\quad {\rm when} \quad
\cos \chi_s \rightarrow
1/3\,.
\label{blowspcuncharged}
\end{equation}
This is equivalent to 
\begin{equation}
\sin^{2} \chi_s = \frac{8}{9}\,.
\label{intschawrlimit0}
\end{equation}
Now,  Eqs.~(\ref{mass0asR}) and (\ref{chi})
allow us to write
\begin{equation}
M = \frac{1}{2}\, R\, \sin^2 \chi_s \, ,
\label{massforbound}
\end{equation}
where $M \equiv m(R)$ is the star's total mass.
Thus, Eqs.~(\ref{intschawrlimit0}) and (\ref{massforbound})
yield
\begin{equation}
\frac{R}{M}  = \frac{9}{4}.
\label{intschawrlimit}
\end{equation}
Equation (\ref{intschawrlimit}) 
is the Schwarzschild limit found in
\cite{incomschwarzschild}.

\subsection{Misner mass bound} 
\label{massbounduncharged}

Following Misner \cite{misner} we can also display a mass bound.  Equation
(\ref{mass0asR}) gives \begin{equation} M =
\frac12\frac{R^3}{R_c^2}\,.  \label{massMasR} \end{equation}
Eliminating $R$ in Eqs.~(\ref{intschawrlimit}) and (\ref{massMasR}),
and noting that $p_c \leq \infty$, one gets the mass bound
\begin{equation} M \leq \frac{1}{2} \,\left( \frac{8}{9}
\right)^{3/2}R_c \, .  \label{mbound} \end{equation} To get a mass we
have to have a density and thus an $R_c$. One can make sense of a
constant density if one takes it as the density at which matter is
almost incompressible and the pressure throughout the star is very
high.  If the fluid is an ideal gas this happens when the particles
have relativistic velocities of the order 1. For fermions this 
happens when the Fermi levels are near the rest mass $m_n$ of the
fermions, neutrons say, while for bosons this means that the gas
temperatures are of the order of the rest mass $m_b$ of the particles.
This gives, for both fermions and bosons, a density of one particle
per cubic Compton wavelength.  I.e., for a particle with mass $m$ and
Compton wavelength $\lambda$ given by $\lambda = \hbar / m$ the
density is $\rho\sim m^4/\hbar^3$.  In the case of a star composed of
neutrons, Misner \cite{misner} obtains
\begin{equation} 
M \leq 1.5
M_{\odot} \, , 
\label{protobbound} 
\end{equation} 
where $M_{\odot}$ is
the Sun's mass.  This bound is similar to the Chandrasekhar limit
$M_{\rm Chandrasekhar} = 1.44 M_{\odot}$, or the Oppenheimer-Volkoff
mass, $M_{\rm OV} \simeq 1 M_{\odot}$, both found for equations of
state different from the one used here and through totally different
means.

\section{The electrically charged 
interior Schwarzschild limit: The small charge case}
\label{smallcharge}

\subsection{Equations: perturbing with a small electric
charge}

\subsubsection{Expansion in the electric charge parameter $\alpha$}

We are going to solve equations 
(\ref{continuidaddamassa}), (\ref{continuidadedacarga}),
and (\ref{tov}), treating the charge 
$q(r)$ as a
small perturbation, thus 
assuming $\alpha$ small. 
To do so, we note that the solutions for
the mass and the charge will be of the form
\begin{eqnarray}
q(r) & =  q_1(r)  \label{charge}\,, \\
m(r) & = m_0(r) +  m_1(r) \,,\label{mass}
\end{eqnarray}
where we are assuming that the non-perturbed charge
is zero $q_0(r)=0$, 
$m_0(r)$ is the mass of the uncharged star given by
Eq.~(\ref{continuidaddamassauncharged}), or
Eq.~(\ref{mass0}), and $q_1(r)$ and $m_1(r)$ are 
the perturbed small charge and mass functions to be
determined. The pressure is also assumed to be given by the
expansion
\begin{equation}
p(r) = p_0(r) + p_1(r)\,, \label{pressure}
\end{equation}
where $p_0$ is the pressure in the uncharged case, given by equation
(\ref{tovuncharged1}), or
(\ref{tovunchargedsol}), and $p_1(r)$ is the perturbation induced in
the pressure when a small charge is considered.  Note that, while the
boundary condition for the non-charged star was simply $p(R) =
p_0(R) = 0$, the boundary condition for the charged star becomes
\begin{equation}
p_0(R) + p_1(R) = 0\,.
\label{cfp1}
\end{equation}

At
this point, it will prove useful to introduce the dimensionless
variable 
\begin{equation}
x=\frac{r}{R_c}\,,
\label{newradial}
\end{equation}
where $R_c$ is the characteristic length defined in
Eq.~(\ref{length}). The expressions for the mass, charge, and pressure
in this new variable are generically
defined as
\begin{equation}
m(x) = \frac{m(r)}{R_c}\,,\quad
q(x) = \frac{q(r)}{R_c}\,,\quad
p(x)=\frac{p(r)}{\rho}\,.
\label{newradialfunctions}
\end{equation}
From Eq.~(\ref{newradial}) we defined $x_s$ 
as the $x$ at the surface, so that
\begin{equation}
x_s=\frac{R}{R_c}\,.
\label{newradials}
\end{equation}
Accordingly, we define
\begin{equation}
m(x_s) = \frac{M}{R_c}\,,\quad
q(x_s) = \frac{Q}{R_c}\,,\quad
p(x_s)=\frac{p(R)}{\rho}\,.
\label{newradialfunctionssurface}
\end{equation}

\subsubsection{Calculation of the perturbed charge
distribution
$q_1$}

Expanding
Eq.~(\ref{continuidadedacarga}) for small $\alpha$, we get
in the $x$ variable that 
\begin{equation}\label{chargesol1}
\frac{dq_1}{dx} = \frac32\,\alpha \frac{x^2}{\sqrt{1- x^2}}\,,
\end{equation}
up to first order in $\alpha$.
Solving the above equation subject to the condition $q_1(0)=0$ 
and expressing the solution in terms of the variable $x$, results in
\begin{equation}\label{chargesol2}
q_1(x) =  \frac{3}{4}\,\alpha\, \left(\arcsin x -x \sqrt{1-x^2}  
\right)\,.
\end{equation}

\subsubsection{Calculation of non-perturbed and perturbed masses}

The unperturbed mass $m_0$ can now be expressed simply as
\begin{equation}\label{m0x}
m_0(x) = \frac{x^3}{2}\,.
\end{equation}
One can also find an expression for $m_1$, namely, 
\begin{equation}\label{pmasssol} 
m_1(x) = \frac{3}{8}\, \alpha^2\,  \left(
3x - x^3 -  3\sqrt{1-x^2} \arcsin x \right)\,.
\end{equation}
Indeed, from Eqs.~(\ref{continuidaddamassa}) and (\ref{mass}), it is
clear that the equation for the perturbed mass $m_1$ is given by
\begin{equation}\label{pmass}
\frac{dm_1}{d x} = \frac{q_1}{x}\frac{dq_1}{dx}\,.
\end{equation}
Inserting Eq.~(\ref{chargesol2})
into Eq.~(\ref{pmass}), we can integrate it using the
boundary condition that the total mass at the center of the star is
$m(0)=0$, which implies that $m_1(0)=0$ since Eq.~(\ref{mass0}), 
or Eq.~(\ref{m0x}), satisfies $m_0(0)=0$. Doing this, we are led to
Eq.~(\ref{pmasssol}).

\subsubsection{Equations for the pressures,
solution for the zeroth order pressure,
and calculation of the
perturbed pressure at 
the star's radius}

\noindent {\it (i) Equations for the pressures}

\noindent
To find the equations 
for the pressures $p_0(x)$ and 
$p_1(x)$, we begin by expressing Eq.~(\ref{tov}) 
for the total pressure in terms of the variable $x$
given in Eq.~(\ref{newradial}),
\begin{eqnarray}\label{tovmodadx}
\frac{\mathrm{d} p}{\mathrm{d} x} =&& - \frac{ \left (1+p(x) \right )
\left ( 3p(x) x/2 + m(x)/x^2 - q^2(x)/x^3 \right) }{1-2m(x)/x +
q^2(x)/x^2} + \nonumber\\
&&+\frac{\alpha q}{x^2 \sqrt{1-2m(x)/x + q^2(x)/x^2}} \, .
\end{eqnarray}
Now we can expanded the right side of the above equation in powers of
$\alpha$ and retain the two lowest terms. By doing so, and using the
expansion (\ref{pressure}) on the left side of Eq.~(\ref{tovmodadx})
and Eqs.~(\ref{charge})-(\ref{mass}) on the right hand side, we can
equate the terms in equal powers of $\alpha$, thus obtaining two
differential equations. The first one, obtained from the 0th power in
$\alpha$ is
\begin{equation}\label{difp0}
\frac{dp_0(x)}{dx} = -\frac{\left (1+p_0(x) \right ) \left
( 3p_0(x)  x/2 + m_0(x)/x^2\right)}{1-2m_0(x)/x},
\end{equation}
which is simply the differential equation for the unperturbed
pressure. The second differential
equation, to first order in $\alpha^2$, is
\begin{eqnarray}\label{difp1}
  \frac{dp_1}{dx}&& = 
\frac{\alpha q_1}{x^2 \, \sqrt{1-2m_0/x}}  
 -\frac{(1+ p_0 )(3 p_0 x/2 + x/2 )f_1}{(1-2 m_0/x)^2}+ \nonumber \\
 &&  - \frac{p_1 (3 p_0 x/2 + x/2)+(1+ p_0)(3p_1 x/2 +
f_2)}{(1-2 m_0/x)},
\end{eqnarray}
which is the differential equation for the perturbed pressure
$p_1$, where, again, to shorten equations we have dropped the dependence of 
variables $p_1$, $p_0$, $m_1$, $m_0$, and $q_1$ on $x$, and we have also 
defined the auxiliary functions
$f_1=f_1(x)$ and $f_2=f_2(x)$ by 
\begin{equation}\label{f1}
f_1(x) = \frac{2m_1(x)}{x} - \frac{q_1^2(x)}{x^2} \, ,
\end{equation}
and
\begin{equation}\label{f2}
f_2(x) = \frac{m_1(x)}{x^2} - \frac{q_1^2(x)}{x^3} \, .
\end{equation}

Ultimately, we 
want to obtain an equation for 
the radius $R$ for which the
central pressure blows up.
From Eq.~~(\ref{pressure}),
the central pressure is $p(0)=p_0(0)+p_1(0)$. 
In the Appendix we show that $p_1(0)$ is 
always finite. So we have to find a solution for 
the 
radius $R$ at which
$p_0(0)$ blows up.

\vskip 0.1cm
\noindent {\it (ii) Solution for the zeroth order pressure}

\noindent
We start by obtaining a solution for $p_0$.
Since the boundary condition has changed
relatively to the uncharged case, 
it is now given by Eq.~(\ref{cfp1}), we cannot use 
a priori the form
(\ref{tovunchargedsol}) for $p_0$. 
We use the solution to
Eq.~(\ref{difp0}) without specifying any boundary condition.
In the variable $x$, this  means
\begin{equation}\label{p0sol}
p_0(x) =  \frac{\sqrt{1-x^2} - A}{3 A - \sqrt{1-x^2}}\,.
\end{equation} where $A$ is an integration constant. To find out what
this constant is, we insert the above equation into Eq.~(\ref{cfp1})
and solve the resulting equation with respect to $A$. After expanding
for the small charge parameter $\alpha$, and so for small $p_1$, we are
led to
\begin{equation}
A = \sqrt{1-x_s^2}  \left(1+2p_1(x_s)\right)\,,
\end{equation}
up to first order in the perturbed 
quantities.
Then,
the expression for $p_0$, analogous to 
Eq.~(\ref{tovunchargedsol}), becomes
\begin{equation}
p_0(x) = \rho \,\frac{\sqrt{1-x^2} - \sqrt{1-x_s^2}
\left(1+2p_1(x_s)\right)}{3 \sqrt{1-x_s^2}
\left(1+2p_1(x_s)\right) - \sqrt{1-x^2} }.
\label{p0final}
\end{equation}
Since $p_0(x)$ depends on $p_1(x_s)$ in the
denominator, we have to find 
$p_1(x_s)$, i.e.,  we have to calculate
the
perturbed pressure at 
the star's radius.

\vskip 0.1cm
\noindent {\it (iii) Calculation of the
perturbed pressure at 
the star's radius}

\noindent
The equation for 
$p_1$, Eq.~(\ref{difp1}),
cannot be solved analytically for all $x$. However, we
are only interested in the value of $p_1$ at the surface of the star.
At this particular radius it is possible to obtain the exact value
of the perturbed pressure without ever solving Eq.~(\ref{difp1}). The
reason for this is the fact that at the star's radius the
pressure $p(x_s)=p_0(x_s)+p_1(x_s)$ 
is zero. Therefore, we can insert the boundary condition
$p(x_s) = 0$ in the exact derivative of the pressure given by
Eq.~(\ref{tovmodadx}) and expand the resulting equation for small
$\alpha$, giving
\begin{eqnarray}\label{difp}
\hspace*{-1.5cm}\frac{dp}{dx}\Big|_{x = x_s} = && -\frac{m_0(x_s)}{x_s^2 (1-2
m_0(x_s)/x_s)} 
- \frac{m_0(x_s)}{1-2 
m_0(x_s)/x_s}\left(\frac{f_1(x_s)}{x_s^2} +\frac{f_2(x_s)}{m_0(x_s)}\right),
\end{eqnarray}
up to first order, and where the auxiliary functions
$f_1(x)$ and $f_2(x)$ are given by Eqs.~(\ref{f1}) and ~(\ref{f2}), 
respectively.
Using the expansion (\ref{pressure}) on the left side of 
Eq.~(\ref{difp}), 
one can clearly see that there is a compatibility condition
which must be physically required, namely that, at the star surface,
the first term on the right side of Eq.~(\ref{difp}) must be equal to
Eq.~(\ref{difp0}) and the second term equal to
Eq.~(\ref{difp1}). Hence, we arrive at the two equations
\begin{equation}\label{p0xs}
\frac{3}{2}\, p_0^2(x_s)+ 2\,p_0(x_s) = 0 
\end{equation}
and
\begin{equation}\label{p1xs}
\frac{3}{2}\,p_1(x_s) p_0(x_s) x_s + p_0(x_s)f_2(x_s) = 0\,.
\end{equation}
These equations give two different solutions, namely, 
$p_0(x_s)=0$ and $p_1(x_s)=0$, or
$p_0(x_s) = -\frac{4}{3}$ and  
$p_1(x_s) =  - \frac{2}{3} x_{s}^{-1} f_2(x_s)$.
This latter solution 
does not satisfy the
boundary condition (\ref{cfp1}) 
so the unique valid solution is given
by 
\begin{equation}
p_0(x_s)=0  \, , \label{sol0}
\end{equation}
and 
\begin{equation}
p_1(x_s)=0 \, . \label{sol1}
\end{equation}

\subsubsection{Equation for the minimum radius}

We see that the central pressure $p_0(0)$ given in Eq.~(\ref{p0final})
is divergent when the following condition holds,
\begin{equation}\label{blowsup} 
3 \sqrt{1-x_s^2} \left ( 1 +
2p_1(x_s) \right ) = 1\,.
\end{equation} 
Expanding it in $\alpha^2$ we arrive to 
\begin{equation}\label{ratio}
\frac{x_s}{m_0(x_s)} = \frac{9}{4}-\frac{9}{8}\,p_1(x_s) \,,
\end{equation}
valid in first order in $\alpha^2$.
Using the expansion 
provided by Eq.~(\ref{mass}), it can 
be shown that to first order in
$\alpha^2$, we have
\begin{equation}\label{ratioabc}
\frac{x_s}{m(x_s)} = \frac{x_s}{m_0(x_s)} -
x_s\,\frac{m_1(x_s)}{m_0^2(x_s)} \,.
\end{equation}
Upon substituting Eq.~(\ref{ratioabc})
into Eq.~(\ref{ratio})
we conclude that
\begin{equation}\label{ratio3_0}
\frac{x_s}{m(x_s)} = \frac{9}{4}-\left(\frac{9}{8}p_1(x_s) +
{x_s}\,\frac{m_1(x_s)}{m_0^2(x_s)}\right) \,.
\end{equation}

Now, the minimum star radius $R$ will not be just $\sqrt{8/9}\, R_c$
but will have corrections of order $\alpha^2$. These corrections will
induce changes of the order $\alpha^4$ in Eq.~(\ref{ratio3_0}).  Thus,
we can set $x_s = \sqrt{8/9}$ in Eq.~(\ref{ratio3_0}), i.e.,
\begin{equation}\label{ratio3}
\frac{x_s}{m(x_s)} = \frac{9}{4}-\left(\frac{9}{8}p_1(x_s) +
 \sqrt{\frac89}\,\frac{m_1(x_s)}{m_0^2(x_s)}\right) \,,
\end{equation}
which is the equation we were looking for.  Since $m_0(x_s)$,
$m_1(x_s)$, and $p_1(x_s)$ can be taken directly from Eq.~(\ref{m0x}),
Eq.~(\ref{pmasssol}), and Eq.~(\ref{sol1}), respectively, we can
proceed to the final result.  Indeed, using Eqs.~(\ref{m0x}),
(\ref{pmasssol}), (\ref{sol1}) in (\ref{ratio3}), we obtain
\begin{equation}\label{ratio4} \frac{x_s}{m_s} = \frac{9}{4} - 1.529\,
\alpha^2\,, \end{equation} up to order $\alpha^2$.

In converting from the variable $x$ back to $r$,
we use
\begin{equation}\label{ratio2}
\frac{x_s}{m(x_s)} = \frac{R}{M}\,.
\end{equation}
In addition, 
it can also be interesting 
to express $\alpha$ in
terms of the total charge $Q$ and mass 
$M$.
The
following relation valid for small $q_1$, or small $\alpha$,
can be found
$\frac{Q}{M}=\frac{ q_1(x_s)}{m_0(x_s) + m_1(x_s) } =
\frac{q_1(x_s)}{m_0(x_s)}$
so that
\begin{equation}
\frac{q_1(x_s)}{m_0(x_s)}=\frac{Q}{M}\,, 
\label{qm000}
\end{equation}		
up to order $q_1$. We can then
express $\alpha$ in terms of the ratio $Q/M$.  Using
Eqs.~(\ref{chargesol2}) and (\ref{m0x}) in Eq.~(\ref{qm000}), and
solving the resulting equation for $\alpha$, we obtain
\begin{equation}
1.641\, \alpha=\frac{Q}{M}\,.
\label{qmasalpha}
\end{equation}

\subsection{The electric interior Schwarzschild limit: The
$R$, $M$ and $Q$ relation for small
charge}

We are now in a position to calculate the desired ratio (\ref{ratio4})
in terms of the quantities $R$, $M$ and $Q$
and find the appropriate relation.
Inserting
Eq.~(\ref{ratio2}) into Eq.~(\ref{ratio4}) we find
\begin{equation}\label{intschwele21}
\frac{R}{M} = \frac{9}{4} - 1.529\, \alpha^2\,,
\end{equation}
which is one form of the interior Schwarzschild limit
for small charge.

Inserting Eq.~(\ref{qmasalpha}) into Eq.~(\ref{intschwele21})
we get
\begin{equation}
\frac{R}{M} = \frac{9}{4} - 0.568 \, \frac{Q^2}{M^2}\,,
\label{intschwele22}
\end{equation}
valid up to order $Q^2/M^2$. This is another 
form of the interior Schwarzschild limit
for small charge.

We can also express the limit in terms of the horizon radius, $r_+$,
for the Reissner-Nordstr\"om metric. The horizon
radius is defined by Eq.~(\ref{externalgravradius}), 
i.e., up to order $Q^2/M^2$ one has, 
$r_+ = M + \sqrt{M^2 - Q^2} = 2M \left(1 -\frac{1}{4}
\frac{Q^2}{M^2} \right)$.
So,
\begin{equation}\label{intschwele23}
\frac{R}{r_+} =  \frac{9}{8}  - 0.003 \frac{Q^2}{M^2}\,, 
\end{equation}
up to order $Q^2/M^2$.
Equation ~(\ref{intschwele23}) is yet another form of the interior
Schwarzschild limit for small charge. 

The electric interior Schwarzschild limit for small charge presented
in various forms in Eqs.~(\ref{intschwele21}), (\ref{intschwele22}),
and (\ref{intschwele23}) is the main result of this work.  All the
three forms of the electric interior Schwarzschild limit for small
charge show that, in comparison with the uncharged case
Eq.~(\ref{intschawrlimit}), the star can be more compact.  In
particular, Eq.~(\ref{intschwele23}) shows that in the charged case
the radius of the star can be a bit nearer its own horizon.

In \cite{arbalemoszanchin} these compact stars were studied
numerically. An $R/M \times Q/M$ relation was given numerically for
$0\leq Q/M\leq 1$. For small charge, $Q/M\ll 1$, one can extract from
the numerical calculations in \cite{arbalemoszanchin} that
$\frac{R}{M} \simeq 2.25 - 0.6\, \frac{Q^2}{M^2}$.  This should be
compared to our analytical calculation valid in first order of $Q/M$,
given here in Eq.~(\ref{intschwele22}), i.e., $\frac{R}{M} =
\frac{9}{4} - 0.568 \, \frac{Q^2}{M^2}$.  It shows that the numerical
code used in \cite{arbalemoszanchin} is compatible with the analytical
calculation.  In that work \cite{arbalemoszanchin} it was also shown
numerically that in the other extreme, namely, $Q/M=1$, one would
obtain a star at its own gravitational radius, $R/M=1$, i.e., an
(extremal) quasiblack hole.

A related theme is the Buchdahl and the Buchdahl-Andr\'easson bounds.
Buchdahl
\cite{buchdahl} by imposing a simple set of assumptions, namely, the
spacetime is spherically symmetric, the star is made of a perfect
fluid, and the density is a nonincreasing function of the radius,
found that the radius to mass relation is $R/M\geq9/4$. Thus the
Schwarzschild limit \cite{incomschwarzschild}, i.e., the limiting
$R/M$ configuration when the central pressure goes to infinity, is an
instance that saturates the Buchdahl bound.  Following the line of
reason of Buchdahl, Andr\'easson
\cite{andreasson_charged} obtained a bound for the
minimum radius of a star using the following energy condition $ p + 2
p_T \leq \rho$ where $p_T$ is the tangential pressure, $p$ is the
radial pressure and $\rho$ is the energy density. This bound, the
Buchdahl-Andr\'easson bound, is given by $\frac{R}{M} \leq
\frac{9}{\left(1 + \sqrt{1 + 3Q^2/R^2} \right)^2}$. Retaining terms in
first order in $Q^2$ one gets $\frac{R}{M} = \frac{9}{4} -
0.667\,\frac{Q^2}{M^2}$.  Thus, our configurations of constant density
and a charge distribution proportional to the energy density having
$\frac{R}{M} = \frac{9}{4} - 0.568\,\frac{Q^2}{M^2}$, see
Eq.~(\ref{intschwele22}), does not saturate the bound. This raises the
question of whether there are other types of charged matter that can
saturate the bound.  One type is thin shells with an appropriate
relation between surface energy density and surface pressure
\cite{andreasson_charged}.  Are there continuous (non-thin-shell)
distribution configurations that saturate the bound?  It seems that,
as the configurations analyzed here, the configurations studied in
\cite{defeliceyu,defelicesiming,annroth} do not saturate the bound.
It remains to be seen if the electrically charged configurations
analyzed in \cite{guilfoyle,lemosezanchin_QBH_pressure} saturate the
bound.  For further study on bounds of electrically charged stars see
\cite{sim,bohmerharko,giulianirothman}.

\subsection{A mass bound}

We can adapt the mass bound from section \ref{massbounduncharged} to
the small charge case.  Indeed, from Eq.~(\ref{mass}) and the
definition $M\equiv m(R)$, we have at the boundary
\begin{equation}
M = m_0(R) + m_1(R) \, .
\label{Msurface}
\end{equation}
Now,  Eq.~(\ref{m0x}) at the boundary can be put in the 
form 
$m_0(R)=  \frac{1}{2} \,R_c \,\sin^3 \chi_s$. 
So, Eq.~(\ref{Msurface}) yields
\begin{equation}
M = \frac{1}{2} R_c \sin^3 \chi_s + m_1(R) \,.
\end{equation}
Since $p_c \leq \infty$, using equation (\ref{ratio}) with $p_1(R)=0$,
one obtains $\sin^2 \chi_s \leq \frac{8}{9}$, and the bound for
the non-perturbed mass $m_0$ is given by $m_0(R) = \frac{1}{2} R_c \sin^3
\chi_s \leq \frac{1}{2}\,\left(\frac{8}{9}\right)^{{3/2}} R_c $.  In
order to obtain the bound for $m_1$, we have to substitute the bound
$\sin^2 \chi_s \leq \frac{8}{9}$ in equation (\ref{pmasssol}). This is
enough since we are only working up to order $\alpha^2$.  Thus, the
mass bound for the small charge case is
\begin{equation}
M \leq \frac{1}{2} \, \left( \frac{8}{9} \,
\right)^{{3/2}} R_c\,\left(1+ 0.679\,\alpha^2\right) \, .
\end{equation}

In the case of a compact star composed of neutrons in the
incompressible state speckled with some charged particles, we obtain
\begin{equation}
M \leq 1.5\, M_{\odot}\, \left(1+ 0.679\,\alpha^2\right)  \, ,
\label{protobboundcharged}
\end{equation}
Comparing equation (\ref{protobboundcharged}) with equation
(\ref{protobbound}) we see that we can attain bigger mass on a charged
star. This is expected since the electrostatic repulsion is opposite
to the gravitational force, which means that we can put more mass on
the star without it collapsing.

\section{Conclusions}\label{sec-conclusion}

In this work we have studied compact stars with a small electric
charge and found the limiting radius to mass relation and the maximum
mass through an analytical approach based on Volkoff and Misner's
method to solve the TOV equation for incompressible matter.  More
specifically, we have analyzed the interior Schwarzschild limit of
spherically symmetric star configurations composed of a fluid with
constant energy density $\rho$ and with a small electrical charge
distribution $\rho_e$ proportional to $\rho$, $\rho_e=\alpha\rho$ with
$\alpha\ll1$.  The exterior spacetime is described by the
Reissner-Nordstr\"om metric.  We have found through our analytical
scheme that due to the electric charge distribution the limiting star
configuration can have more mass and a smaller radius relatively to
the limiting star with zero charge. This is expected since the
electric charge distribution has a repulsive effect, adding to the
pressure as a force that withstands the star.  For stars containing
some type of dark matter in their interior there is the possibility
that they possess a small but non-negligible electric charge, in which
case our analytical formula is opportune and can be confronted
quantitatively with observational data.

\section*{Acknowledgments}
We thank FCT-Portugal for financial support through Project
No.~PEst-OE/FIS/UI0099/2014.
JPSL and VTZ thank CAPES for the 
Programa Pesquisador Visitante Especial
Project
No.~88887.068694/2014-00.

\appendix \section
{Behaviour of the perturbed pressure for small radius}
\label{appendixA}

Our goal is to compute the limit of infinite central pressure $p(r=0)$
in this charged case, so it is important to analyze the behaviour of
$p_1$ for small radius $x$, $x \sim 0$, which also means $\chi \sim
0$, and $r \sim 0$. We start by obtaining the perturbed charge, $q_1$,
for small radius expanding equation (\ref{chargesol2}), which gives
\begin{equation}
q_1(x) \sim \alpha \frac{x^3}{2}\, , \label{chargesmallr} 
\end{equation}
and the perturbed mass, $m_1$, from equation (\ref{pmasssol})
\begin{equation} 
m_1(x) \sim \alpha^2 \frac{3}{20}x^5\, . \label{pmasssmallr} 
\end{equation}
The non-perturbed pressure, $p_0$, given 
by equation (\ref{p0final}) becomes
\begin{equation}
p_0 (x) \sim p_0(x=0) - \frac{\sqrt{1-x_s^2}}{3 \sqrt{1-x_s^2}- 1} x^2
\,.  
\label{p0smallr}
\end{equation}
Using equation (\ref{difp1}) for the perturbed pressure, $p_1$,
together with equations (\ref{chargesmallr}), (\ref{pmasssmallr}) and
(\ref{p0final}) it gives
\begin{equation}
\frac{\mathrm{d} p_1}{\mathrm{d} x} \sim -p_1 \left (3
p_0(x=0) + 2 \right) x + \frac12\,\alpha^2 \, x \,. 
\label{difp1smallr}
\end{equation}
Integrating equation (\ref{difp1smallr}) one obtains
\begin{eqnarray}
\hspace*{-1.7cm} p_1(x) \sim && \frac{\alpha^2}{6p_0(x=0)+4}
+\left( p_1(x=0)- 
\frac{\alpha^2}{6p_0(x=0)+4} \right) e^{-(3p_0(x=0)+2)x^2/2} \,. 
\label{p1smallr}
\end{eqnarray}
So the unique way that the central pressure $p(r)=p_0(r) +
p_1(r)$ blows up 
is $p_0(r=0)$ blowing up.




\end{document}